\documentclass[12pt,english]{article}
\usepackage{mathptmx}

\usepackage[T1]{fontenc}
\usepackage[latin9]{inputenc}
\usepackage{amssymb}
\usepackage{babel}
\begin{document}

\title{Quantum Entanglement and Generalized Uncertainty Relations}

\author{Harris Varghese, M. Raveendranadhan\thanks{{\footnotesize{}ravicusat@cusat.ac.in : raviskvc@gmail.com }\protect \\
{\footnotesize{}harrisvarghese@gmail.com}}{\footnotesize{}{}}\\
 {\footnotesize{}{DEPARTMENT OF PHYSICS}}\\
 {\footnotesize{}{COCHIN UNIVERSITY OF SCIENCE AND TECHNOLOGY}}\\
 {\footnotesize{}{KOCHI, KERALA, INDIA}}\\
 {\footnotesize{}{PIN-682002}}}

\maketitle
The Robertson's formulation of the uncertainty relation is the most
widely accepted form of the Heisenberg uncertainty relation (HUR).
It gets modified when we consider it for entangled particles. But
this formulation does not consider the measurement process itself.
There are reformulation of the uncertainty relations called the generalized
uncertainty relations by including the measurement process into the
uncertainty relation. HUR gets modified for the case of entangled
particles. Here it is shown that the limit of the Generalized Uncertainty
Relation (GUR) also reduces for entangled particles. So, GUR also
shows similar trend as that of the HUR for entangled particles. Also
as entanglement increases, the uncertainty reduces and measurement
becomes more precise. 

\section{Introduction}

Robertson's relation is the most widely accepted form of the uncertainty
relation. For two observables $A\;and\;B$ on a state $\psi$ it is
defined as 
\begin{equation}
\sigma(A,\psi)\sigma(B,\psi)\geq\frac{|\langle\psi|[A,B]|\psi\rangle|}{2}\label{eq:1}
\end{equation}
where $\sigma$ is the standard deviation, also known as the uncertainty
in the measurement of $\hat{A},$ is defined as $\sigma(A)=\langle(A-\langle A\rangle)^{2}\rangle^{\frac{1}{2}}$
. Left side of equation (\ref{eq:1}) usually known as the uncertainty
in the simultaneous measurement of $\hat{A}$ and $\hat{B}$ on a
particle with state $\psi.$ If there are more particles, the state
$\psi$ is a product state or entangled state of the particles. Using
quantum covariance function\cite{QCF} Rigolin \cite{Rigolin16} has
explained how the entanglement between particles affects the uncertainty
relation. It is shown that entanglement reduces the uncertainty in
the simultaneous measurement of $\hat{A}$ and $\hat{B}.$ It is also
shown that right side of equation (\ref{eq:1}) tends to zero with
large number of entangled particles.

The Robertson's relation does not consider the measurement process,
it only depends on the state of the system. In realistic situation
measurement will introduce noise in the measured value disturbance
to other quantities.

Measurement noise and the corresponding disturbance on the system
are not included Robertson's uncertainty relation. Also there is experimental
violation of the usual uncertainty relations\cite{expviolatnofHUR}.
Ozawa\cite{ozawaHur} proposed a universal Generalized Uncertainty
Relations (GUR) which are the reformulations of the usual uncertainty
relation by considering the measurement process also. It was further
modified by Fujikawa \cite{FujikwaGUR}.

In this paper we are studying how the generalized uncertainty relations
are affected by the build-up of entanglement on the system. We consider
(GUR) by Fujikawa in presence two particles. The discussion can extended
to $n$ entangled particles. 

\section{Noise and Disturbance}

The root-mean-square noise $\epsilon(A)$ of a measuring device measuring
$A$ is given as\cite{ozawanoiseop} 
\begin{equation}
\epsilon(A)^{2}=\langle(M^{out}-A^{in})^{2}\rangle=\langle\psi\otimes\zeta|(U^{\dagger}(I\otimes M)U-(A\otimes I))^{2}|\psi\otimes\zeta\rangle
\end{equation}
where $|\psi\rangle$ is the state of the system of particles and
$|\zeta\rangle$ is the state of the measuring device (probe) and
the unitary operator $U$ on $\psi\otimes\zeta$ gives the time evolution
of the composite system (system+probe) during their interaction. $M$
is the probe observable to be detected from the state after the measuring
interaction. From now on $\langle\dots\rangle$ stands for $\langle\psi\otimes\zeta|\dots|\psi\otimes\zeta\rangle$.
The root-mean-square disturbance $\eta(B)$ is defined as\cite{ozawanoiseop}
\begin{equation}
\eta(B)^{2}=\langle(B^{out}-B^{in})^{2}\rangle=\langle\psi\otimes\zeta|(U^{\dagger}(B\otimes I)U-(B\otimes I))^{2}|\psi\otimes\zeta\rangle\label{eq:3}
\end{equation}
It is assumed that the measuring device measuring $B$ is noiseless.
so 
\[
U^{\dagger}(B\otimes I)U=U^{\dagger}(I\otimes M_{B})U
\]
and then it is only affected by the measurement of $A$ during the
interaction. 

GUR due to Ozawa is
\begin{equation}
\epsilon(A)\eta(B)+\epsilon(A)\sigma(B)+\sigma(A)\eta(B)\ge\frac{\left|\left\langle \psi\left|[A,B]\right|\psi\right\rangle \right|}{2}
\end{equation}
where $\epsilon$ is the noise in the measurement of $A$ and $\eta$
is the corresponding disturbance on $B$. It was further modified
by Fujikawa \cite{FujikwaGUR} as 
\begin{equation}
\epsilon(A)\eta(B)+\epsilon(A)\sigma(B)+\sigma(A)\eta(B)+\sigma(A)\sigma(B)\ge\frac{\left|\left\langle \psi\left|[A,B]\right|\psi\right\rangle \right|}{2}\label{eq:5}
\end{equation}
In the absence of noise and disturbance, above relation reduces to
Robertson's elation (\ref{eq:1}). To investigate many particle effect
on Fujikawa relation we consider a system of two distinguishable particles. 

For a two particle system let us define position operator 
\begin{equation}
Q(1,2)=Q_{1}\otimes I_{2}+I_{1}\otimes Q_{2}\label{eq:6}
\end{equation}
for particles at $q_{1}$ and $q_{2}$ and measurement operator 
\begin{equation}
M_{Q(1,2)}=M_{Q_{1}}\otimes I_{2}+I_{1}\otimes M_{Q_{2}}\label{eq:7}
\end{equation}
Where $M_{Q_{1}}$ is measurement operator representing measurement
$Q_{1}$ on particle 1 and $M_{Q_{2}}$ is measurement operator representing
measurement $Q_{2}$ on particle 2. Corresponding to this, noise on
the measurement of $Q(1,2)$ is
\begin{eqnarray*}
\epsilon(Q(1,2))^{2} & = & \langle(U^{\dagger}(I\otimes M_{Q(1,2)})U-(Q(1,2)\otimes I))^{2}\rangle\\
 & = & \langle(U^{\dagger}(I\otimes M_{Q_{1}}\otimes I_{2}+I\otimes I_{1}\otimes M_{Q_{2}})U\\
 &  & -(Q_{1}\otimes I_{2}\otimes I+I_{1}\otimes Q_{2}\otimes I))^{2}\rangle\\
 & = & \langle(U^{\dagger}(I\otimes M_{Q_{1}}\otimes I_{2})U-(Q_{1}\otimes I_{2}\otimes I))^{2}\rangle+\\
 &  & \langle(U^{\dagger}(I\otimes I_{1}\otimes M_{Q_{2}})U-(I_{1}\otimes Q_{2}\otimes I)){}^{2}\rangle+\\
 &  & 2\langle U^{\dagger}(I\otimes M_{Q_{1}}\otimes I_{2})(I\otimes I_{1}\otimes M_{Q_{2}})U\rangle+\\
 &  & 2\langle(Q_{1}\otimes I_{2}\otimes I)(I_{1}\otimes Q_{2}\otimes I)\rangle-\\
 &  & 2\langle U^{\dagger}(I\otimes M_{Q_{1}}\otimes I_{2})U(I_{1}\otimes Q_{2}\otimes I)\rangle-\\
 &  & 2\langle U^{\dagger}(I\otimes I_{1}\otimes M_{Q_{2}})U(Q_{1}\otimes I_{2}\otimes I)\rangle
\end{eqnarray*}
Since the operators are commuting, we could rearrange and obtain it
as 
\begin{eqnarray}
\epsilon(Q(1,2))^{2} & = & \epsilon(Q{}_{1})^{2}+\epsilon(Q{}_{2})^{2}+2[(\langle U^{\dagger}(I\otimes M_{Q_{1}}\otimes I_{2})U\rangle-\nonumber \\
 &  & \langle(Q_{1}\otimes I_{2}\otimes I)\rangle)(\langle U^{\dagger}(I\otimes I_{1}\otimes M_{Q_{2}})U\rangle-\langle(I_{1}\otimes Q_{2}\otimes I)\rangle)\label{eqn 8}
\end{eqnarray}
For the simultaneous measurement of position of both the particles
using a symmetric experimental setup, the noise is not the sum of
noise of measurement on the first particle and the second particle,
$\epsilon(Q(1,2))^{2}\ne\epsilon(Q{}_{1})^{2}+\epsilon(Q{}_{2})^{2}$.
There is an additional term. If the measurement on the first particle
is noiseless, then $\epsilon(Q(1,2))^{2}=\epsilon(Q{}_{2})^{2}$.
Similarly, if the measurement on the second particle is noiseless,
then we have $\epsilon(Q(1,2))^{2}=\epsilon(Q{}_{1})^{2}$. So, the
additional term should contain the noise of the measurement on the
first particle and also on the second particle. Now we define 
\begin{equation}
\epsilon(Q(1,2))=\langle U^{\dagger}(I\otimes M_{Q(1,2)})U-(Q(1,2)\otimes I)\rangle
\end{equation}
By definition, note that 
\begin{equation}
\epsilon(Q(1))^{2}\neq\epsilon(Q(1))\epsilon(Q(1))
\end{equation}
Here $\epsilon(Q(1))^{2}$ is the noise in the measurement of $Q(1)$
in presence of $Q(2)$. Then we could write the equation (\ref{eqn 8})
as 
\begin{equation}
\epsilon(Q(1,2))^{2}=\epsilon(Q{}_{1})^{2}+\epsilon(Q{}_{2})^{2}+2\epsilon(Q{}_{1})\epsilon(Q{}_{2})
\end{equation}
Evidently when $\epsilon(Q_{1})=0\;(\epsilon(Q_{2})=0)$ we get $\epsilon(Q(1,2))=\epsilon(Q_{1}\;(\epsilon(Q(1,2))=\epsilon(Q_{2})$.
Now we could similarly consider the disturbance caused by this position
measurements on the system. We may assume that disturbance caused
is to momentum. As in the case of position, we define the momentum
operator as 
\[
P(1,2)=P_{1}\otimes I_{2}+I_{1}\otimes P_{2}
\]
By taking $B=P\left(1,2\right)$ in equation (\ref{eq:3}) we get
root-mean-square disturbance $\eta(P\left(1,2\right))$ as
\begin{equation}
\eta(P(1,2))^{2}=\eta(P{}_{1})^{2}+\eta(P{}_{2})^{2}+2\eta(P{}_{1})\eta(P{}_{2})\label{eq:12}
\end{equation}

\section{Entanglement and GUR}

Now assume that using symmetric experimental techniques, we made the
noise and disturbance on the first particle is equal to that of the
second particle. That is when $Q_{1}=Q_{2}$ and $P_{1}=P_{2}$, we
would have 
\begin{eqnarray}
\epsilon(Q)^{2} & = & 2\epsilon(Q_{1})^{2}+2[\epsilon(Q_{1})]^{2}\\
\eta(P)^{2} & = & 2\eta(P_{1})^{2}+2[\eta(P_{1})]^{2}
\end{eqnarray}
Consider that our two particle system is prepared in such a way that
the two particles are entangled. When measurement occurs, due to the
interaction with the measuring devices the entanglement between the
particles get destroyed. So while calculating the noise and disturbance
due to measurement on these particles we could treat them as separable
states. Then for the separable states 
\begin{equation}
\epsilon(Q_{1})^{2}=[\epsilon(Q_{1})]^{2}\;\;\;and\;\;\;\eta(P_{1})^{2}=[\eta(P_{1})]^{2}\label{eq:16a}
\end{equation}
and then we get 
\begin{equation}
\epsilon(Q)=2\epsilon(Q)_{1}\;\;\;and\;\;\;\eta(P)=2\eta(P)_{1}\label{eqn 17}
\end{equation}

The uncertainty in position and momentum for the preparation is given
by the standard deviations of position and momentum in the state.
The standard deviations depend only on the state of the system and
it is independent of the property of the measuring apparatus or measurement.
That is measurement does not affect the standard deviation. So the
standard deviation for two entangled particles is given as

\[
\sigma_{Q}=\sqrt{\langle Q(1,2)^{2}\rangle-\langle Q(1,2)\rangle^{2}}
\]

Now substituting equation \ref{eq:6}

\[
\sigma_{Q}=\sqrt{(\Delta Q_{1})^{2}+(\Delta Q_{2})^{2}+2\langle Q_{1}\otimes Q_{2}\rangle-2\langle Q_{1}\rangle\langle Q_{2}\rangle}
\]

Using the quantum covariant function\cite{QCF}, for entangled particles
we could arrive at\cite{Rigolin16}

\begin{equation}
\sigma_{Q}=\sqrt{(\Delta Q_{1})^{2}+(\Delta Q_{2})^{2}+(\Delta Q_{1})^{2}+(\Delta Q_{2})^{2})}
\end{equation}
If the preparation of our system is carried out in a way that, the
deviation in position and momentum is same for the first and second
particles. Then we would get 
\begin{equation}
\sigma_{Q}=2\Delta Q_{1}=2\sigma_{Q_{1}}\label{eqn 19}
\end{equation}
Using similar assumption 
\begin{equation}
\sigma_{P}=2\Delta P_{1}=2\sigma_{P_{1}}\label{eqn 20}
\end{equation}
Now for our two particle system we know that \\
 
\begin{equation}
Q(1,2),P(1,2)]=[Q_{1},P_{1}]+[Q_{2},P_{2}]=2i\hbar\label{eqn 21}
\end{equation}
Then by substituting equations (\ref{eqn 17}), (\ref{eqn 19}), (\ref{eqn 20})
and (\ref{eqn 21}) on to the generalized uncertainty relation by
Ozawa, we could get 
\begin{eqnarray}
\epsilon(Q)\eta(P)+\epsilon(Q)\sigma(P)+\sigma(Q)\eta(P) & \ge & \frac{|\langle\psi|[Q,P]|\psi\rangle|}{2}\nonumber \\
2\epsilon_{Q_{1}}2\eta_{P_{1}}+2\epsilon_{Q_{1}}2\sigma_{P_{1}}+2\sigma_{Q_{1}}2\eta_{P_{1}} & \ge & \frac{2\hbar}{2}\nonumber \\
4(\epsilon_{Q_{1}}\eta_{P_{1}}+\epsilon_{Q_{1}}\sigma_{P_{1}}+\sigma_{Q_{1}}\eta_{P_{1}}) & \ge & \hbar\nonumber \\
\epsilon_{Q_{1}}\eta_{P_{1}}+\epsilon_{Q_{1}}\sigma_{P_{1}}+\sigma_{Q_{1}}\eta_{P_{1}} & \ge & \frac{\hbar}{4}
\end{eqnarray}
The usual limit in the uncertainty is $\frac{\hbar}{2}$, when the
particles gets entangled the limit reduces to half of the traditional
one. \\
 Similarly for the Generalized Uncertainty Relation by Fujikawa we
have 
\begin{eqnarray}
\epsilon(Q)\eta(P)+\epsilon(Q)\sigma(P)+\sigma(Q)\eta(P)+\sigma(Q)\sigma(P) & \ge & |\langle\psi|[Q,P]|\psi\rangle|\nonumber \\
\epsilon_{Q_{1}}\eta_{P_{1}}+\epsilon_{Q_{1}}\sigma_{P_{1}}+\sigma_{Q_{1}}\eta_{P_{1}}+\sigma_{Q_{1}}\sigma_{P_{1}} & \ge & \frac{\hbar}{2}
\end{eqnarray}
This is also only half of the traditional limit. Thus entanglement
causes the limit of the uncertainty to become smaller. Similar to
that of the Robertson's relation, both generalized uncertainty relations
by Ozawa and Fujikawa also reduces for the case of entangled particles.
It can be easily shown that as the number of entangled particles increases
the uncertainty reduces even further. That is, as entanglement increases
the uncertainty reduces even further. It causes our measurements to
become more precise. So when we consider more entanglement between
the particles, the system is becoming more classical. 

\section{Conclusions}

The limit of generalized uncertainty relation reduces when we are
observing an entangled system. Our measurement becomes more precise
in the case of entangled particles, which is similar to that of Robertson's
relation. Uncertainty in the system reduces as entanglement develops
in the system. Also as the number of entangled particles increases
the uncertainty reduces further, it results the system to become more
classical. So it is predicted that entanglement holds the key to the
transition from quantum realm to classical realm.  \bibliographystyle{unsrt}
\bibliography{BIB}

\begin{thebibliography}{1}

\bibitem{QCF}
A.~C. de~la Torre, P.~Catuogno, and S.~Ferrando.
\newblock Uncertainty and nonseparability.
\newblock {\em Foundations of Physics Letters}, 2:235--243, 1989.

\bibitem{Rigolin16}
Rigolin G.
\newblock Entanglement and the uncertainty principle.
\newblock {\em Communications in Theoretical Physics}, 66:201, 2016.

\bibitem{expviolatnofHUR}
Jacqueline Erhart, Stephan Sponar, Georg Sulyok, Gerald Badurek, Masanao Ozawa,
  and Yuji Hasegawa.
\newblock Experimental demonstration of a universally valid error disturbance
  uncertainty relation in spin measurements.
\newblock {\em Nature Physics}, 8:185--189, March 2012.

\bibitem{ozawaHur}
Masanao Ozawa.
\newblock Universally valid reformulation of the heisenberg uncertainty
  principle on noise and disturbance in measurement.
\newblock {\em Physical Review A}, 67:042105 1--6, 2003.

\bibitem{FujikwaGUR}
Kazuo Fujikawa.
\newblock Universally valid heisenberg uncertainty relation.
\newblock {\em Physical Review A}, 85:062117 1--3, 2012.

\bibitem{ozawanoiseop}
Masanao Ozawa.
\newblock Universal uncertainty principle in the measurement operator
  formalism.
\newblock {\em Journal of Optics B: Quantum and Semiclassical Optics},
  7:S672--S681, 2005.

\end{thebibliography}
 
\end{document}